# Physics Beyond the Standard Model

*R. Rosenfeld*
ICTP South American Institute for Fundamental Research & Instituto de Física Teórica
Universidade Estadual Paulista, São Paulo, Brazil

**Abstract**
In these three lectures I review the need to go beyond the Standard Glashow-Weinberg-Salam Model and discuss some of the approaches that are explored in this direction.

## 1   Introduction

In these lectures I took a somewhat different approach to introduce the possibility of new physics beyond the Standard Model (SM). Usually one finds in most lectures discussions about well motivated models proposed to solve the most pressing issue of the SM: the hierarchy problem. Hence one usually presents the three extensions that tackles this issue: supersymmetric models, composite models and models with compact extra dimensions.

At the expense of being less motivated, I decided to not introduce these grand schemes and instead be more model-independent and making only minimal modifications to the SM, restricting myself to the scalar sector. These modifications already gives a rough idea of some the consequences of expanding the SM.

In this manner after a discussion of the hierarchy problem I started with the assumption that no new particles will be directly found at accelerators, possibly due to the large mass scales involved. In this case new physics would manifest itself in the form of new higher dimensional operators induced by integrating out these heavy new degrees of freedom and suppressed by the scale of new physics. This is the spirit of the Effective Theory approach. Afterwards I introduce minimal modifications in the scalar sector, which already produces modified Higgs couplings and a dark matter candidate. In the final lecture the traditional models for physics beyond the SM, such as SUSY, Composite Higgs and Extra Dimensions are briefly touched upon.

I must say right away that in my opinion a full-fledged write-up of these lectures, which still need much improvement, is not essential at this point. Hence these notes are intended as just a brief guide to what was discussed. The slides of my lectures, as well as for the other lectures of the School can be found at:
`http://physicschool.web.cern.ch/PhysicSchool/CLASHEP/CLASHEP2015/programme.html`
There is some overlap with the excellent lectures by Christophe Grojean on Higgs Physics in this School.

In addition, there are several lectures on Physics Beyond the Standard Model, among which I list:

– Lectures by Alex Pomarol at the 2010 CERN School [1];
– Lectures by Eduardo Pontón at the 2012 TASI School [2];
– Lectures by Joe Lykken at the 2009 CERN School [3];
– Lectures by Tony Gherghetta at the 2009 TASI School [4]

Therefore in the following I will present a simple sketch of my lectures pointing to some references where more details can be found.

## 2 First lecture

In this lecture I started by recalling the astounding success of the SM, exemplified by the muon and electron magnetic moments, the Z boson line shape and the number of neutrinos, precision electroweak measurements at LEP, measurements of cross sections for SM processes at the LHC and, finally, the historical discovery of a Higgs boson in 2012. I also briefly mentioned the recent measurements of Higgs couplings which are in agreement with SM predictions within experimental errors.

Next the usual shortcomings of the SM were mentioned, noticing that it does not explain several issues:

- the 19 free parameters necessary to explain the observed phenomena (flavor problem, absence of strong CP violation, etc);
- origin of electroweak symmetry breaking;
- neutrino masses;
- dark matter;
- origin of the matter-antimatter asymmetry;
- dark energy;
- inflation;
- gravity.

In addition, we discussed some of the more conceptual problems related to the scalar sector of the SM: perturbative unitarity, triviality, vacuum stability and, especially, the hierarchy or naturalness problem. I recommend the review of Giudice [5] for a very lucid explanation of the naturalness problem.

In particular, the hierarchy problem arises from the absence of a symmetry that could protect the Higgs mass from receiving dangerous quadratic quantum corrections. It has played a major role in the development of models beyond the SM. Favorite models such as Supersymmetry, Composite Higgs and Compact Extra Dimensions have naturalness as their main motivation. All solutions to the naturalness/hierarchy problem lead to new physics at scales not much above the electroweak scale. The discovery of a light Higgs boson and the absence of any new particles or deviations of couplings at the LHC have put theories motivated by the naturalness principle under stress. It has been called "the LHC battle for naturalness"[1].

If the mass scale of new physics is beyond the reach of the LHC their main effects can be parameterized in an effective lagrangian that includes higher dimensional terms involving only SM fields that respect the known SM symmetries. This approach is very general and can be called agnostic in the sense that it does not depend on the details of the underlying model. The number of terms is finite but can be large. At dimension-6, there are 59 terms that can be added to the SM lagrangian [6].

Under certain assumptions, the absence of new physics results in constraints on the energy scale associate to it. However, one must keep in mind that a given observable may be sensitive to a combination of dimension-6 operators and on the other hand each dimension-6 operator may affect more than one observable. Typically, barring spurious cancellations among different operators one obtains bounds from LEP electroweak precision measurements from LEP1 alone that the scale of new physics $\Lambda$ should be larger than $O(10 \text{ TeV})$. LEP2 and LHC bounds are less restrictive, roughly $\Lambda > 300$ GeV.

The least experimentally constrained operators are the ones involving the third generation, such as top quark dipole operators. These are also expected to receive contributions if new physics couples dominantly to the third generation, which is the case for several SM extensions.

I also briefly mentioned that an Effective Lagrangian approach can also be used to parameterize the interaction of a dark matter sector to the SM [7] and this has been used in experimental searches [8].

---

[1] See https://indico.cern.ch/event/290373/

I concluded the first lecture remarking that Effective Lagrangians are an indirect and agnostic way to study new physics. One might say it represents the lamp post approach – trying to find new phenomena hiding in error bars. It is difficult to derive firm conclusion in this approach: bounds usually depend on combinations of Wilson coefficients and energy scale(s) of new physics. In the end we will only be convinced of new physics by direct evidence!

## 3 Second lecture

The Higgs boson may be the first fundamental scalar particle found in Nature. It is conceivable that there are more scalar particles out there in a "hidden sector". They may communicate to us only via the Higgs: the Higgs acts like a portal between the SM and this new sector.

In the second lecture I discuss the simplest extension of the SM: the addition of a singlet real scalar field which we denote by $S$ interacting only with the Higgs doublet via a renormalizable term in the potential. This is not motivated by any grand principle such as naturalness but illustrates some phenomena common to some more complete extensions.

Already in this simple extension one has 2 choices (with different phenomenology) in writing the potential: to allow or not for $S$ to have a vacuum expectation value (vev).

### 3.1  $\langle S \rangle \neq 0$

If $S$ has a nonzero vev it can mix with the Higgs boson, generating two mass eigenstates that we call $H_1$ and $H_2$, and identifying $H_1$ with the 125 GeV scalar found at the LHC. There are three additional parameters with respect to the SM, related to the S mass, self-coupling and coupling to the Higgs doublet.

In this model there are some phenomenological consequences:

- all Higgs couplings are reduced by a common factor of $\cos\theta$, where $\theta$ is a mixing angle. Hence all Higgs partial widths are reduced by $\cos^2\theta$ with respect to the SM value;
- couplings of the second Higgs to gauge bosons and fermion are the same of a SM Higgs reduced by $\sin\theta$;
- there are new processes (depending on the mass of $H_2$):
  $H_2 \to H_1 H_1$ if $m_{H_2} > 250$ GeV and $H_1 \to H_2 H_2$ if $m_{H_2} < 62.5$ GeV.

There are many bounds in this model coming from:

- perturbativity of couplings;
- vacuum stability (potential bounded from below);
- EW precision measurements: modified couplings, new loop contributions from $H_2$;
- LEP direct searches (low mass $H_2$);
- LHC direct searches (high mass $H_2$);
- Higgs couplings at LHC ($H_1 \to \gamma\gamma$, $4f$) (modification of widths due to mixing, possible new contribution to $H_1$ width for light $H_2$);
- partial unitarization.

The bounds on the parameters of this simple extension can be found in [9, 10].

The possibility of a resonant di-Higgs production through $pp \to H_2 \to H_1 H_1$ for a heavy $H_2$ is interesting since the SM cross section for double Higgs production is very small. In particular, if $H_2$ is very heavy the decays of $H_1$ will be boosted and jet substructure techniques can be used to search for this process in the final state with 4 $b$ quarks [11]. This has been searched for by CMS [12] and ATLAS [13]. There was also a search in the rarer $bb\gamma\gamma$ channel by CMS [14].

## 3.2 ⟨S⟩ = 0

It is possible that there is an unbroken $Z_2$ symmetry in the scalar sector under which $S \leftrightarrow -S$. This symmetry forbids the field $S$ to develop a vacuum expectation value. It also makes the $S$ boson stable. Hence this model is arguably the simplest extension of the SM with a dark matter candidate [15]. In fact, since $S$ has self-interactions this is a model of self-interacting dark matter. The $Z_2$ symmetry also forbids a mixing between the new scalar field with the Higgs field. However, the Higgs boson can decay into two dark matter particles, $H \to SS$ leading to invisible Higgs decays [16, 17].

A term such as $\lambda_{HS} S^2 H^2$ controls:

- $SS \leftrightarrow SMSM$ (annihilation to SM particles, that determine DM relic abundance);
- $SN \to SN$ (elastic scattering off nucleons, that determine DM direct detection);
- $H \to SS$ (invisible Higgs decay).

A term such as $S^4$ controls DM self-interactions.

This model has few free parameters and is already severely constrained. For a recent analysis see [18].

One comment about these models is that it introduces a new physical scale: the mass of the particle $S$. This in turn generates a hierarchy problem since the $S$ particle induces a new contribution to the Higgs boson mass. Again, naturalness implies that the $S$ particle can not be too heavy, typically:

$$M_S^2 < \frac{16\pi^2}{\lambda_{HS}} M_H^2$$

Many extensions of the SM build on this simple class of models, just adding more scalar fields: complex singlet, 2-Higgs doublets (inert or active, SUSY), Higgs triplets, etc.

Although the vacuum stability issue at high energies can be ameliorated, the simple model discussed here was not built to avoid the naturalness problem, which is arguably the guiding principle to BSM. The next lecture is about models that were motivated by the hierarchy/naturalness problem.

## 4 Third lecture

In this lecture I begin by discussing the naturalness problem for the electron mass when taking into account the electron self-energy due to its own electrical field. The self-energy contribution is actually divergent and this problem was eventually solved in Quantum Field Theory, where one knows that the electron mass is protected by chiral symmetry - the self-energy of the electron computed in QFT is proportional to the electron mass itself.

This type of mechanism led 't Hooft to conjecture the following "dogma" in 1980 that he called naturalness: "at any energy scale $\mu$ a physical parameter or set of physical parameters $\alpha_i(\mu)$ is allowed to be very small only if the replacement $\alpha_i(\mu) = 0$ would increase the symmetry of the system". For example, setting the mass of the electron to zero restores chiral symmetry in QED. This dogma was promoted into a principle in later years.

Setting the Higgs mass to zero in the SM does not increase any symmetry. There is no natural reason for why the Higgs boson should be light. In fact, there are quantum quadratic contributions to the Higgs boson mass that makes it sensitive to very high energy scales. Requiring that in the SM the contributions to the Higgs boson mass from a scale $\Lambda$ is smaller that the Higgs boson mass itself requires $\Lambda < 600$ GeV. But LHC has ruled out New Physics at this scale. Thus the LHC has shown that SM is not natural.

This is the "naturalness" motivation to go BSM: find a mechanism that can explain why $M_H << \Lambda$ for large values of the cut-off representing a physical scale where New Physics should show up.

This requires either a new symmetry to protect the Higgs mass or a mechanism to lower the cut-off. In the first category we may list:

- Supersymmetry (cancellation of quadratic divergences);
- Shift symmetry (Higgs as a pseudo-Nambu-Goldstone boson);
- Conformal symmetry (Higgs as a dilaton);

whereas the latter case includes

- Flat extra-dimensions (Large Extra Dimensions, Universal Extra Dimensions);
- Warped Extra Dimensions (Randall-Sundrum models)

I cannot go into the details about these different alternatives in these Proceedings. There is a vast literature on these naturalness-motivated BSM. It suffices to say that even in the models mentioned above the absence of New Physics at the LHC is calling into question the naturalness principle. The second run of the LHC at 13 TeV that has just started should bring very important information in the coming years.

## 5 Acknowledgements

I would like to thank Nick Ellis, Martijn Mulders, Kate Ross and especially Edgar Carrera for the organization of CLASHEP, the invitation to lecture and for the hospitality in Ecuador.